\documentclass[12pt]{elsart}
\usepackage{amsmath}
\usepackage{amssymb}
\usepackage{amsfonts}
\usepackage{graphics}
\usepackage{latexsym}
\usepackage{epsfig}

\graphicspath{plots}
\newcommand{\eq}{\begin{equation}}
\newcommand{\en}{\end{equation}}
\newcommand{\eqa}{\begin{eqnarray}}
\newcommand{\ena}{\end{eqnarray}}

\begin{document}
\vspace*{-1.9cm}
\begin{flushleft}
{\normalsize DESY 01-035} \hfill\\
{\normalsize LTH-495} \hfill\\
{\normalsize March 2001} 
\end{flushleft}
\vspace{0.2cm}
\begin{frontmatter}
  \title{\bf Determination of $\mathbf{\Lambda_{\overline{MS}}}$ from quenched
   and $\mathbf{N_f = 2}$ dynamical QCD}

\vspace*{-0.55cm}

  \author[EPCC]{S. Booth},
  \author[RE]{M. G\"ockeler},
  \author[NI]{R. Horsley},
  \author[LI]{A.C. Irving},
  \author[ED]{B. Joo}~\footnote{Present address: Department of Physics, 
  Columbia University, New York, NY 10027, USA},
  \author[ED]{S. Pickles}~\footnote{Present address: Computer Services for 
  Academic Research CSAR, University of Manchester, Manchester M13 9PL, UK},
  \author[NI]{D. Pleiter},
  \author[RE]{P.E.L. Rakow},
  \author[NI,DE]{G. Schierholz},  
  \author[ED]{Z. Sroczynski}~\footnote{Present address: Fachbereich Physik, 
  Universit\"at Wuppertal, D-42097 Wuppertal, Germany},
  \author[ZU]{H. St\"uben}
 
\vspace{0.15cm}

{-- QCDSF--UKQCD {\it Collaboration} --} 

\vspace{0.25cm}
  
  \address[EPCC]{Edinburgh Parallel Computing Center EPCC,\\ 
   University of Edinburgh, Edinburgh EH9 3JZ, UK}
  \address[RE]{Institut f\"ur Theoretische Physik, Universit\"at Regensburg,\\
   D-93040 Regensburg, Germany}
  \address[NI]{John von Neumann-Institut f\"ur Computing NIC,\\
   Deutsches Elektronen-Synchrotron DESY,
   D-15735 Zeuthen, Germany} 
  \address[LI]{Theoretical Physics Division, Department of Mathematical 
   Sciences,\\ University of Liverpool, Liverpool L69 3BX, UK}
  \address[ED]{Department of Physics and Astronomy,\\ 
   University of Edinburgh, Edinburgh EH9 3JZ, UK}
  \address[DE]{Deutsches Elektronen-Synchrotron DESY,
   D-22603 Hamburg, Germany}
  \address[ZU]{Konrad-Zuse-Zentrum f\"ur Informationstechnik Berlin,
   D-14195 Berlin, Germany}

  \date{ }  
%

\begin{abstract}

The scale parameter $\Lambda_{\overline{MS}}$ is computed on the lattice in
the quenched approximation and for $N_f = 2$ flavors of light dynamical
quarks. The dynamical calculation is done with non-perturbatively $O(a)$ 
improved Wilson fermions. In the continuum limit we obtain 
$\Lambda_{\overline{MS}}^{N_f=0} = 243(1)(10)\,\mbox{MeV}$ and
$\Lambda_{\overline{MS}}^{N_f=2} = 217(16)(11)\,\mbox{MeV}$, respectively.
\end{abstract}


\end{frontmatter}

\pagebreak

\section{Introduction}

The $\Lambda$ parameter sets the scale in QCD. In the chiral limit it is 
the only parameter of the theory, and hence it is a quantity of fundamental 
interest. 
It is defined by the running of the strong coupling constant 
$\alpha_s$~\cite{PDG} at high energies where non-perturbative effects 
are supposed to become small. 
Lattice gauge theory provides a means of determining $\alpha_s$ 
directly from low-energy quantities. In this letter we shall compute 
$\Lambda$ on the lattice in the quenched approximation as well as for 
$N_f = 2$ species of degenerate dynamical quarks.

Previous lattice calculations have employed a variety of methods to
compute the strong coupling constant, in quenched and unquenched 
simulations. For reviews see~\cite{weisz,shigemitsu}. The scale 
parameter $\Lambda$
has been extracted from the heavy-quark potential~\cite{ukqcd,bali,Edwards}, 
the quark-gluon vertex~\cite{skullerud}, the three-gluon 
vertex~\cite{alles,boucaud}, from the spectrum of heavy 
quarkonia~\cite{aida,aoki,wingate,davies,sesam}, and by means of
finite-size-scaling methods~\cite{alpha}.  

We determine $\Lambda$ from the average plaquette and the force parameter 
$r_0$~\cite{Sommer}. Both quantities are widely computed in lattice 
simulations. In the 
quenched case we have many data points over a wide range of 
couplings at our disposal already, and in the dynamical case we expect to 
accumulate more points in the near future.
At present $r_0$ is the best known lattice quantity, at least in full QCD. 
It can easily be replaced with more physical scale parameters like hadron 
masses or $f_\pi$ when the respective data become more accurate.

\section{Method}
  
The calculations are done with the standard gauge field action
\begin{equation}
S_G = \beta \sum_x \mbox{Re}\,\frac{1}{3}\mbox{Tr} U_{\Box}(x),
\end{equation}
and, in the dynamical case, with non-perturbatively $O(a)$ improved Wilson 
fermions~\cite{SW}
\begin{equation}
S_F = S^{(0)}_F - \frac{\rm i}{2} \kappa_{\rm sea}\, g\, c_{SW} a a^4 
\sum_x \bar{\psi}(x)\sigma_{\mu\nu}F_{\mu\nu}\psi(x),
\label{action}
\end{equation}
where $S^{(0)}_F$ is the original Wilson action and $\beta = 6/g^2$. If 
the improvement coefficient $c_{SW}$ is appropriately
chosen, this action removes all $O(a)$ errors from on-shell quantities. A
non-perturbative evaluation of this function leads to the
parameterization~\cite{Jansen}
\begin{equation}
c_{SW} = \frac{1-0.454 g^2 - 0.175 g^4 - 0.012 g^6 + 0.045 g^8}{1 - 0.720 g^2}
\label{csw}
\end{equation}
for $N_f = 2$ flavors, which is valid for $\beta \geq 5.2$.

The running of the coupling is described by the $\beta$ function defined by
\begin{equation}
\mu \frac{ \partial g_{\mathcal{S}}(\mu) }{\partial \mu} = 
\beta^{\mathcal{S}} ( g_{\mathcal{S}} (\mu) ),
\label{rge}
\end{equation}
where $\mathcal{S}$ is any mass independent renormalization scheme. The 
perturbative expansion of the $\beta$ function reads
\begin{equation}
\beta^{\mathcal{S}}(g_{\mathcal{S}}) = -g_{\mathcal{S}}^3 \, \big( b_0 
+ b_1 g_{\mathcal{S}}^2 + b_2^{\mathcal{S}} g_{\mathcal{S}}^4 
+ b_3^{\mathcal{S}} g_{\mathcal{S}}^6 +\cdots\; \big). 
\end{equation} 
The first two coefficients are universal, 
\begin{equation}
\begin{split}
b_0 &= \frac{1}{(4\pi)^2} \big(11 -\frac{2}{3}N_f \big) , \\
b_1 &= \frac{1}{(4\pi)^4} \big( 102
  - \frac{38}{3}N_f \big) , 
\label{bs} 
\end{split}
\end{equation} 
while the others are scheme dependent. The renormalization group equation
(\ref{rge}) can be exactly solved:
\begin{equation}
\frac{\mu}{\Lambda_{\mathcal{S}}} =
\big( b_0 g_{\mathcal{S}}^2 \big)^{\frac{b_1}{2 b_0^2} } \,
\exp \Big( \frac{1}{ 2 b_0 g_{\mathcal{S}}^2 }  
+ \int_0^{g_{\mathcal{S}}} {\rm d}\xi \; 
\big( \frac{1}{\beta^{\mathcal{S}}(\xi)} + \frac{1}{b_0 \xi^3 } 
- \frac{b_1}{b_0^2 \xi} \big) \Big), 
\label{soln} 
\end{equation} 
where the scale parameter $\Lambda$ appears as the integration constant. In 
the $\overline{MS}$ scheme the $\beta$ function is known to four 
loops~\cite{Rit}:
\begin{equation}
\begin{split}
b_2^{\overline{MS}} &= \frac{1}{(4 \pi)^6}\big(\frac{2857}{2}
- N_f \frac{5033}{18} + N_f^2 \frac{325}{54}\big) , \\[1em]
b_3^{\overline{MS}} &= 
\frac{1}{(4 \pi)^8}\Big(\frac{149753}{6} + 3564\, \zeta_3 
- N_f \big(\frac{1078361}{162} + \frac{6508}{27}\, \zeta_3\big) \\
&\hspace{1.8cm}+ N_f^2 \big(\frac{50065}{162} + \frac{6472}{81}\, \zeta_3\big)
+ N_f^3 \frac{1093}{729}\Big) ,
\label{bms} 
\end{split}
\end{equation} 
where $\zeta_3 = 1.20206 \cdots$ is Riemann's zeta function.

\begin{figure}[tbp]
\vspace*{-1.0cm}
  \begin{center}
    \epsfig{file=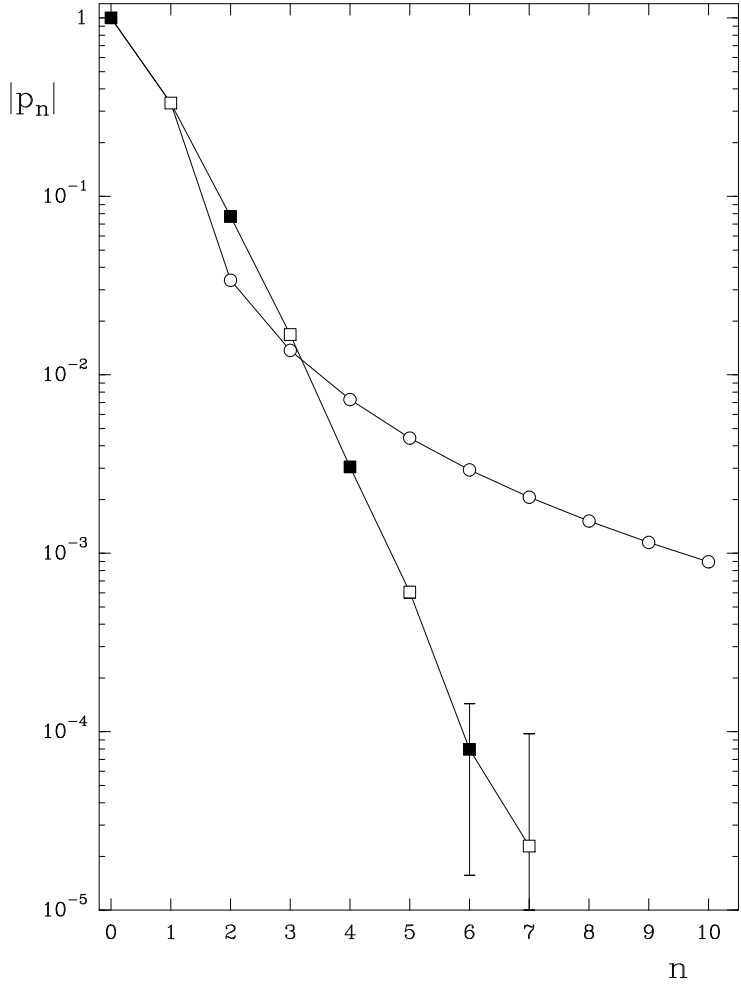,width=10.5cm}
\vspace*{-1.0cm}
\caption{The expansion coefficients for the quenched plaquette, $1 - P =
\sum_{n=1}^\infty p_n x^n$, for $x=g^2$ (\lower0.85pt\hbox{\Large 
$\circ$}) and $x=g_\Box^2$ ($\square$ and $\blacksquare$), respectively.
Open (solid) symbols refer to positive (negative) numbers. The bare
expansion coefficients (\lower0.85pt\hbox{\Large $\circ$}) are taken 
from~\cite{Parma}. Note that the 
boosted coefficients are not only much smaller and rapidly decreasing, but 
also alternating in sign.}
    \label{fig1}
  \end{center}
\end{figure}

In this paper we are concerned with three different schemes. In the continuum
we use the $\overline{MS}$ scheme. On the lattice we consider the bare coupling
$g(a)$ and the boosted coupling $g_\Box(a)$. The latter
is given by
\begin{equation}
g^2_\Box(a) = \frac{g^2(a)}{P} ,
\label{gbox}
\end{equation}
where $P= (1/3)\: \langle\mbox{Tr} U_\Box\rangle \equiv u_0^4$ is the average 
plaquette value. The widespread opinion is that the perturbative expansion in
$g_\Box$ converges more rapidly than the expansion in the bare 
coupling~\cite{Lepage}. 
Indeed, a comparison between the expansion coefficients in the two cases for
the quenched plaquette shown in Fig.~1 supports this belief, as even for low
orders the new series has oscillating coefficients. The conversion 
from the bare coupling to the $\overline{MS}$ scheme has the form
\begin{equation} 
\frac{1}{g_{\overline{MS}}^2(\mu)} = 
\frac{1}{g^2(a)} +  2 b_0 \ln a \mu - t_1
+ ( 2 b_1 \ln a \mu - t_2 ) \, g^2(a) 
+ O( g^4 \ln^2 a \mu ) .
\label{tdef} 
\end{equation} 
Writing
\begin{equation}
\frac{1}{g_\Box^2}  = \frac{1}{g^2} - p_1 -p_2 g^2 - p_3 g^4 - \cdots ,
\label{plaq}
\end{equation} 
we obtain the relation between the boosted coupling and the $\overline{MS}$
coupling
\begin{equation} 
\begin{split}
\frac{1}{g_{\overline{MS}}^2(\mu)} &= 
\frac{1}{g_\Box^2(a)} +  2 b_0 \ln a \mu - t_1 + p_1 \\ 
&+ ( 2 b_1 \ln a \mu - t_2 + p_2) \, g^2_\Box(a) + O( g^4 \ln^2 a \mu ) . 
\label{boxms} 
\end{split}
\end{equation} 
By differentiating eq.~(\ref{boxms}) we can find the $\beta$ function 
coefficient, $b_2^\Box$, for the
boosted coupling
\begin{equation}
b_2^\Box = b_2^{\overline{MS}} + b_0 (p_2-t_2) - b_1(p_1-t_1).
\end{equation}

The one-loop coefficients $p_1$ and $t_1$ are given by
\begin{align}
p_1 &= \frac{1}{3} , \\[0.3em]
t_1 &= 0.4682013 - N_f \big( 0.0066960 - 0.0050467\, c_{SW} 
     +  0.0298435\, c_{SW}^2 \nonumber  \\ 
\label{t_dyn}
    &\hspace{0.4cm}+ am ( -0.0272837 + 0.0223503\, c_{SW} 
     - 0.0070667\, c_{SW}^2 )\big) , 
\end{align}
where $m$ is the quark mass.
The quenched coefficients are taken from \cite{Luscher}, whereas 
the fermionic contribution, including the improvement term 
proportional to $am$, is computed in the Appendix. The pure
Wilson ($c_{SW} = 0$) result agrees with \cite{Alles}, and the $m=0$  
result agrees with the number quoted in~\cite{Bode}.
The $m=0$ two-loop coefficients $p_2$ and $t_2$ are given by
\begin{align}
\label{twop}
p_2 &= 0.0339110 - N_f \frac{8}{3} (0.0006929 - 0.0000202\, c_{SW} 
     + 0.0005962\, c_{SW}^2) , \\[0.3em] 
t_2 &= 0.0556675 -  N_f (0.002600 + 0.000155\, c_{SW} 
     - 0.012834\, c_{SW}^2 \nonumber \\
\label{twot}
    &\hspace{0.4cm}- 0.000474\, c_{SW}^3 - 0.000104\, c_{SW}^4) .
\end{align}
The two-loop $am$ term is not known. 
The quenched coefficients can be found in \cite{Luscher}, whereas the
fermionic contribution to $p_2$ is given in \cite{Christine}, and $t_2$ has 
been computed in \cite{Haris}. 

Combining these terms gives 
\begin{equation} 
\frac{1}{g_{\overline{MS}}^2(\mu)} =
 \frac{1}{g_\Box^2(a)} +  2 b_0 \ln a \mu - t_1^\Box 
 + ( 2 b_1 \ln a \mu - t_2^\Box) \, g^2_\Box(a) + \cdots 
 \label{boxcon}
\end{equation} 
with
\begin{align}
t_1^\Box &= 0.1348680 - N_f \big(0.0066960 - 0.0050467\, c_{SW}  
          +  0.0298435\, c_{SW}^2 \nonumber \\ 
         &\hspace{0.4cm}+ am ( -0.0272837 + 0.0223503\, c_{SW} 
          - 0.0070667\, c_{SW}^2 )\big) , \\[0.3em]
t_2^\Box &= 0.0217565 - N_f ( 0.000753 + 0.000209 \, c_{SW}
          -  0.014424 \,c_{SW}^2 \nonumber \\
\label{t2box}
         &\hspace{0.4cm}- 0.000474 \,c_{SW}^3 - 0.000104 \,c_{SW}^4 ) . 
\end{align} 
Note that $t_1^\Box \ll t_1$, so that the series converting $g_\Box$ to 
$g_{\overline{MS}}$, eq.~(\ref{boxcon}), is better behaved than 
the original series converting bare $g$ to $g_{\overline{MS}}$, 
eq.~(\ref{tdef}). We can improve the convergence of the series further by 
re-expressing it in terms of the tadpole improved coefficients~\cite{Lepage}
\begin{equation}
\begin{split}
{\widetilde c}_{SW} &\equiv c_{SW} u_0^3 , \\
a{\widetilde m} &\equiv am / u_0 . 
\end{split}
\label{tad}
\end{equation}
We then obtain
\begin{equation} 
\frac{1}{g_{\overline{MS}}^2(\mu)} =
\frac{1}{g_\Box^2(a)} +  2 b_0 \ln a \mu - {\widetilde t}_1^\Box 
 + ( 2 b_1 \ln a \mu - {\widetilde t}_2^\Box) \, g^2_\Box(a) + \cdots 
 \label{tildser}
\end{equation} 
with
\begin{align}
{\widetilde t}_1^\Box &= 0.1348680 - N_f \big( 0.0066960 - 0.0050467 \,
{\widetilde c}_{SW} +  0.0298435 \,{\widetilde c}_{SW}^2 \nonumber \\ 
                      &\hspace{0.4cm}+ a{\widetilde m}( -0.0272837
+ 0.0223503 \,{\widetilde c}_{SW} - 0.0070667 \,{\widetilde c}_{SW}^2 ) 
\big) ,  \\[0.3em]
{\widetilde t}_2^\Box &= 0.0217565 - N_f ( 0.000753
- 0.001053 \, {\widetilde c}_{SW}
+ 0.000498 \,{\widetilde c}_{SW}^2 \nonumber \\
\label{tildbox}
                      &\hspace{0.4cm}- 0.000474 \,{\widetilde c}_{SW}^3
- 0.000104 \,{\widetilde c}_{SW}^4 ) . 
\end{align} 
Changing $t_1^\Box$ to ${\widetilde t}_1^\Box$ is simply a matter
of replacing every $c_{SW}$ by  ${\widetilde c}_{SW}$ and every
$m$ by  ${\widetilde m}$, but the change in $t_2^\Box$ is not
so simple, because the coefficients of the $c_{SW}$ and $c_{SW}^2$ terms
change. We see that tadpole improvement is successful in reducing the
two-loop fermionic contribution: the largest coefficient in the fermionic
part of $t_2^\Box$ was $0.01442\dots$, in ${\widetilde t}_2^\Box$ it
is $0.00105\dots$\:. 

We are still free to choose the scale $\mu$ in eq.~(\ref{tildser}). A good 
value to help eq.~(\ref{tildser}) to converge rapidly is to choose $\mu$ so 
that the $O(g^0)$ term vanishes. Therefore we choose the scale so that
\begin{equation} 
a \mu = \exp \left( \frac{{\widetilde t}_1^\Box}{2 b_0} \right) .
\label{scale} 
\end{equation} 
In the quenched case this gives $\mu = 2.63/a$, while for $N_f = 2$ dynamical 
fermions $\mu \approx 1.4/a$. Substituting this scale into eq.~(\ref{tildser}),
we obtain the relationship
\begin{equation}
g^2_{\overline{MS}}(\mu ) = g^2_\Box(a) +
 \left( {\widetilde t}_2^\Box - \frac{b_1}{b_0} {\widetilde t}_1^\Box 
 \right) g^6_\Box(a) + O(g^8) ,
\label{convert} 
\end{equation} 
which agrees with \cite{Luscher} in the quenched case. 

The calculation of $\Lambda_{\overline{MS}}$ proceeds in four steps. First
we compute the average plaquette. From eq.~(\ref{gbox}) we then obtain
$g_\Box$. In the second step we use eq.~(\ref{convert}) to calculate
$g_{\overline{MS}}$ at the scale $\mu$. Putting this value of 
$g_{\overline{MS}}$ into eq.~(\ref{soln}) gives us 
$\mu/\Lambda_{\overline{MS}}$. Finally we use
the conversion factor eq.~(\ref{scale}) to turn this into a value for
$a \Lambda_{\overline{MS}}$.
To convert our results to a physical scale, we use the force parameter
$r_0$.

\begin{figure}[b]
\begin{center}
\vspace*{0.2cm}
\epsfig{file=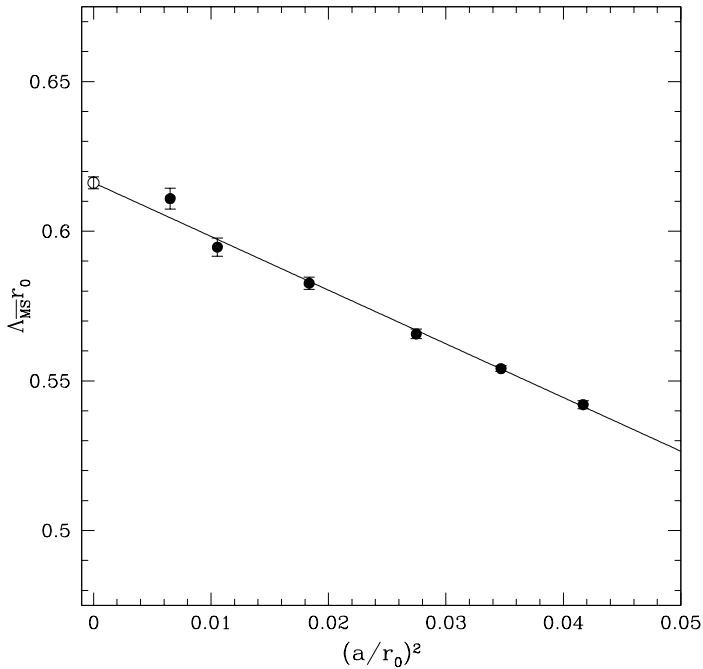,width=9.5cm}
\vspace*{-0.2cm}
\caption{The quenched scale parameter $\Lambda_{\overline{MS}}\,r_0$ against 
$(a/r_0)^2$ together with the continuum value (\ref{qres}). The coupling 
ranges from $\beta = 5.95$ to $6.57$. The line is a linear extrapolation 
to the continuum limit.}
\vspace{0.5cm}
\label{fig2}
\end{center}
\end{figure}

\section{Results}

\noindent
$\underline{N_f=0}$
\vspace*{0.5cm}

Let us begin with the quenched case. The plaquette values 
are taken from QCDSF's quenched simulations~\cite{QCDSF,Dirk}, except at 
$\beta = 6.57$ where the plaquette value is obtained by interpolation.
The $r_0$ values are taken from~\cite{Edwards,Guagnelli}. 
In Fig.~2 we plot $\Lambda_{\overline{MS}}\, r_0$ against
$(a/r_0)^2$. The corresponding numbers are given in Table~1. 
One expects discretization errors of $O(a^2)$. Indeed, the data
points lie on a straight line, allowing a linear extrapolation to the continuum
limit. This gives
\begin{equation}
\Lambda_{\overline{MS}}^{N_f=0} r_0 = 0.616(2)(25) ,
\label{qres}
\end{equation}
where the first error is purely statistical, while the second one is an 
estimate of the systematic error. The latter is derived by assuming that the 
higher-order contributions in eq.~(\ref{convert}) are about 20\% of the 
$O(g^6)$ term. Using $r_0 = 0.5 \: \mbox{fm}$, we find
\begin{equation}
\Lambda_{\overline{MS}}^{N_f=0} = 243(1)(10)\:\mbox{MeV}. 
\label{qqres}
\end{equation}
Our result agrees very well with the outcome of previous lattice
calculations~\cite{boucaud,alpha}. 

It should be noted that $r_0$ is a phenomenological quantity, though a very
robust one, which introduces an additional systematic error. By comparing the 
results of various potential models we estimate the error to be less than 5\%.
Taking the $\rho$ mass to set the scale 
gives~\cite{Dirk} $r_0 = 0.52(2)\:\mbox{fm}$, which is consistent with the 
value used in eq.~(\ref{qqres}). 

\begin{table}[t]
\begin{center}
\vspace{0.2cm}
\begin{tabular}{|c|l|l|l|} \hline
$\beta$ & \multicolumn{1}{|c|}{$P$} & \multicolumn{1}{|c|}{$(a/r_0)^2$}&
\multicolumn{1}{|c|}{$\Lambda_{\overline{MS}}r_0$}\\
\hline
 5.95 & 0.588006(20) & 0.04168(20) & 0.5420(13) \\
 6.00 & 0.593679(8)  & 0.03469(12) & 0.5541(9)  \\
 6.07 & 0.601099(18) & 0.02748(16) & 0.5659(16) \\
 6.20 & 0.613633(2)  & 0.01836(13) & 0.5826(21) \\
 6.40 & 0.630633(4)  & 0.01054(11) & 0.5947(31) \\
 6.57 & 0.6434(2)    & 0.00653(7)  & 0.6109(35) \\
\hline
\end{tabular}\vspace{0.7cm}
\caption{The quenched $\Lambda_{\overline{MS}}\,r_0$ values, together
with $(a/r_0)^2$ and the plaquette $P$.}  
\end{center}
\vspace{0.4cm}
\end{table}

\vspace*{0.3cm}
\noindent
$\underline{N_f=2}$
\vspace*{0.5cm}

In the dynamical case we use combined results from the QCDSF and UKQCD
collaborations~\cite{Hinnerk,Alan}. The gauge field configurations were 
obtained using the
standard Hybrid Monte Carlo algorithm with the non-perturbatively $O(a)$
improved action (\ref{action}). Details of the extraction of $r_0/a$ are
given in~\cite{UKQCD2}.
For the quark mass $m$ we take the Ward identity mass. We compute this mass in 
the same way~\cite{Dirk2} as in the quenched case~\cite{QCDSF}, with the 
improvement coefficient $c_A$ taken from tadpole improved perturbation 
theory. The relevant parameters and results are given in Table 2. The number
of gauge field configurations varies from $O(500)$ on the $16^3 32$ lattices 
to $O(300)$ on the $24^3 48$ lattice. 

\begin{table}[t]
\begin{center}
\vspace{0.2cm}
\begin{tabular}{|c|c|c|c|l|l|l|l|} \hline
$\beta$ & $\kappa_{\mbox{sea}}$ & V  & $c_{SW}$ &\multicolumn{1}{|c|}{$P$} 
&\multicolumn{1}{|c|}{$r_0/a$}& \multicolumn{1}{|c|}{$am$}&
\multicolumn{1}{|c|}{$\Lambda_{\overline{MS}}\,r_0$}\\
\hline
$\!5.20\! $ & $\!0.1355\!$ & $\!16^3 32\!$ &$\!2.0171\!$
 & $\!0.536294(9)\!$ & $\!5.041(40)\!$ &$\!0.02364(16)\!$ &$\!0.4744(38)\!$ \\
$\!5.20\! $ & $\!0.1350\!$ & $\!16^3 32\!$ &$\shortparallel$
 & $\!0.533676(9)\!$ & $\!4.754(40)\!$ &$\!0.04586(19)\!$ &$\!0.4593(39)\!$ \\
$\!5.25\! $ & $\!0.1352\!$ & $\!16^3 32\!$ &$\!1.9603\!$
 & $\!0.541135(24)\!$& $\!5.137(49)\!$ &$\!0.04268(17)\!$ &$\!0.4666(45)\!$ \\
$\!5.26\! $ & $\!0.1345\!$ & $\!16^3 32\!$ &$\!1.9497\!$
 & $\!0.539732(9)\!$ & $\!4.708(52)\!$ &$\!0.07196(20)\!$ &$\!0.4348(48)\!$ \\
$\!5.29\! $ & $\!0.1355\!$ & $\!24^3 48\!$ &$\!1.9192\!$
 & $\!0.547081(26)\!$& $\!5.62(9)\!$  &$\!0.03495(12)\!$  &$\!0.4834(77)\!$ \\
$\!5.29\! $ & $\!0.1350\!$ & $\!16^3 32\!$ &$\shortparallel$
 & $\!0.545520(29)\!$& $\!5.26(7)\!$  &$\!0.05348(19)\!$  &$\!0.4601(61)\!$ \\
$\!5.29\! $ & $\!0.1340\!$ & $\!16^3 32\!$ &$\shortparallel$
 & $\!0.542410(9)\!$ & $\!4.813(45)\!$ &$\! 0.09272(29)\!$  &$\!0.4355(41)\!$\\
\hline
\end{tabular}
\vspace{0.7cm}
\caption{The dynamical $\Lambda_{\overline{MS}}\,r_0$ values, together
with $r_0/a$, $P$ and the quark masses. The improvement 
coefficient $c_{SW}$ was taken from eq.~(\ref{csw}).}
\end{center}
\vspace{0.4cm}
\end{table}

\begin{figure}[b]
\begin{center}
\vspace{0.2cm}
\epsfig{file=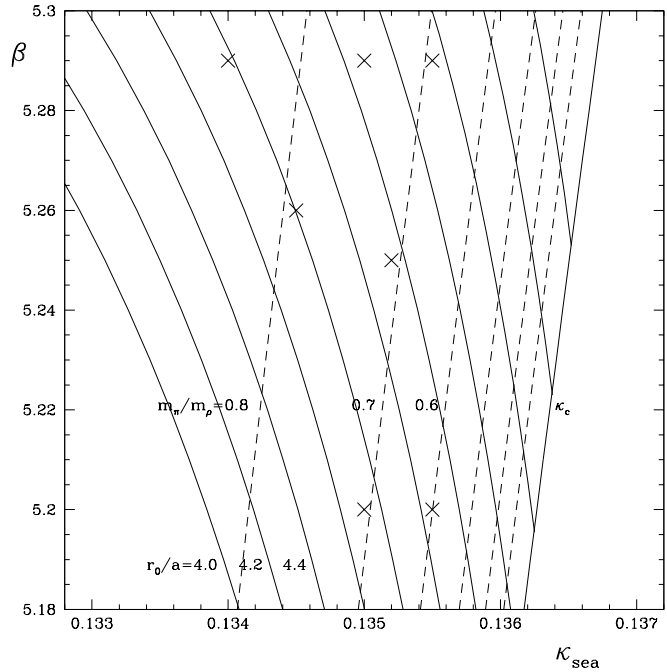,width=9.5cm}
\vspace*{-0.2cm}
\caption{Lines of constant $r_0/a$ (full lines), from 4.0 (left) to 6.0 
(right), and constant $m_\pi/m_\rho$ (dashed lines), from 0.8 to 0.3, 
together with $\kappa_c$ and
the parameters ($\times$) of our simulations (up to now) in the 
($\kappa_{\rm sea}$, $\beta$) plane. The curves are from a fit to the 
renormalization group and chiral perturbation theory, respectively.}
\vspace{0.5cm}
\label{fig3}
\end{center}
\end{figure}

As we are working at finite quark mass, we have to perform an extrapolation to
the chiral limit. In Fig.~3 we show the parameter values of our simulations
together with lines of constant $r_0/a$ and $m_\pi/m_\rho$. This gives an
impression of how far our simulations are from the chiral and continuum limits.

The value that interests us is $\Lambda_{\overline{MS}}$ at $m \rightarrow 0$
and $a \rightarrow 0$. Given the fact that our action has discretization 
errors of $O(a^2)$ only, at least as $m \rightarrow 0$, we expect the 
following small-$a$ behavior:
$\Lambda_{\overline{MS}}(a) = \Lambda_{\overline{MS}}(a=0)(1+b_\Lambda am + 
O((a/r_0)^2) + O((am)^2))$, where $\Lambda_{\overline{MS}}(a=0)$ is not 
supposed 
to depend on $m$ anymore. Similarly, we expect to find $r_0(a) = r_0(a=0)
(1+b_r am + O((a/r_0)^2) + O((am)^2))$, with the difference that $r_0(a=0)$ 
may still depend on $m$: $r_0(a=0) = r_0(a=0,m=0)(1+c_r mr_0 + O((mr_0)^2))$. 
Putting everything together, we then arrive at the following parameterization 
of $\Lambda_{\overline{MS}}\, r_0$ for small $a, m$:
\begin{equation}
\Lambda_{\overline{MS}}\, r_0 = A (1 + B am)(1 + C mr_0) + D (a/r_0)^2,
\label{extra}
\end{equation}
where we have neglected terms of $O((mr_0)^2)$. Effectively 
$\Lambda_{\overline{MS}}\, r_0$ can be written as a function of $mr_0$ and 
$a/r_0$. 

We do not know $c_A$ non-perturbatively. It turns out though that the final 
result is not affected by a small adjustment of $c_A$, for this changes 
all masses by a common factor, within the statistical errors, and hence 
amounts to a rescaling of the fit parameters $B$ and $C$ only.

Let us now turn to the fit and extrapolation of our data. In the fit we 
assume that $\Lambda_{\overline{MS}}\,r_0$, $am$ and $r_0/a$ are uncorrelated.
We find that the ansatz (\ref{extra}) fits the data very well ($\chi^2 = 
3.1$). The parameters $B$ and $C$ are strongly correlated though, indicating
that it does not matter whether we are using $am$ or $mr_0$ as the chiral
extrapolation variable. Indeed, fixing $B = 0$ gives the same result for $A$
and an almost identical value of $\chi^2$. 
To justify our ansatz (\ref{extra}), we subtract the 
mass dependence from the measured values of $\Lambda$ to obtain 
$\Lambda_{\overline{MS}}\, r_0(m=0) \equiv
\Lambda_{\overline{MS}}\, r_0 - A (B am + C mr_0 + B C am\,mr_0)$.
A plot of $\Lambda_{\overline{MS}}\, r_0(m=0)$ against $(a/r_0)^2$ should 
then collapse all data points onto the single line $A + D (a/r_0)^2$.
In the presence of significant higher-order terms not covered by our ansatz 
we would, on the other hand, expect 
to see the data deviate from that line. Similarly, a plot of 
$\Lambda_{\overline{MS}}\, r_0(a=0) \equiv (\Lambda_{\overline{MS}}\, 
r_0 - D (a/r_0)^2)/(1 + B am)$ against $mr_0$ should collapse the data onto 
the line $A (1 + C mr_0)$. This 
is what we have plotted in Figs.~4 and 5. We see that it does indeed bring all
data points onto one line. We also see that the deviations from the line are
probably not statistically significant, so adding any extra term to the fit,
like $(m r_0)^2$, the deviation from the line is just going to give a fit to 
the noise. In fact, we have experimented with higher-order polynomials in $a$
and $m$. In all cases we found the same result in the chiral and continuum 
limit within the statistical error. 
Note that the slope of the line in Fig.~4 is very similar to the 
slope of the corresponding quenched line in Fig.~2.

In the chiral and continuum limit our fit gives 
\begin{equation}
\Lambda_{\overline{MS}}^{N_f=2} r_0 = 0.549(39)(28) .
\label{contd}
\end{equation}
The first error is purely statistical, while the second one is an 
estimate of the systematic error, where we again have assumed that the 
higher-order contributions in eq.~(\ref{convert}) are about 20\% of the 
$O(g^6)$ term. Using $r_0 = 0.5 \: \mbox{fm}$, this gives
\begin{equation}
\Lambda_{\overline{MS}}^{N_f=2} = 217(16)(11)\:\mbox{MeV} .
\end{equation}
A preliminary computation of the mass spectrum yields
$r_0 = 0.50(7)$ fm, if we take the $\rho$ mass to set the scale, 
in agreement with the phenomenological value used.   

\clearpage 
\begin{figure}[ht]
\begin{center}
\vspace*{-0.8cm}
\epsfig{file=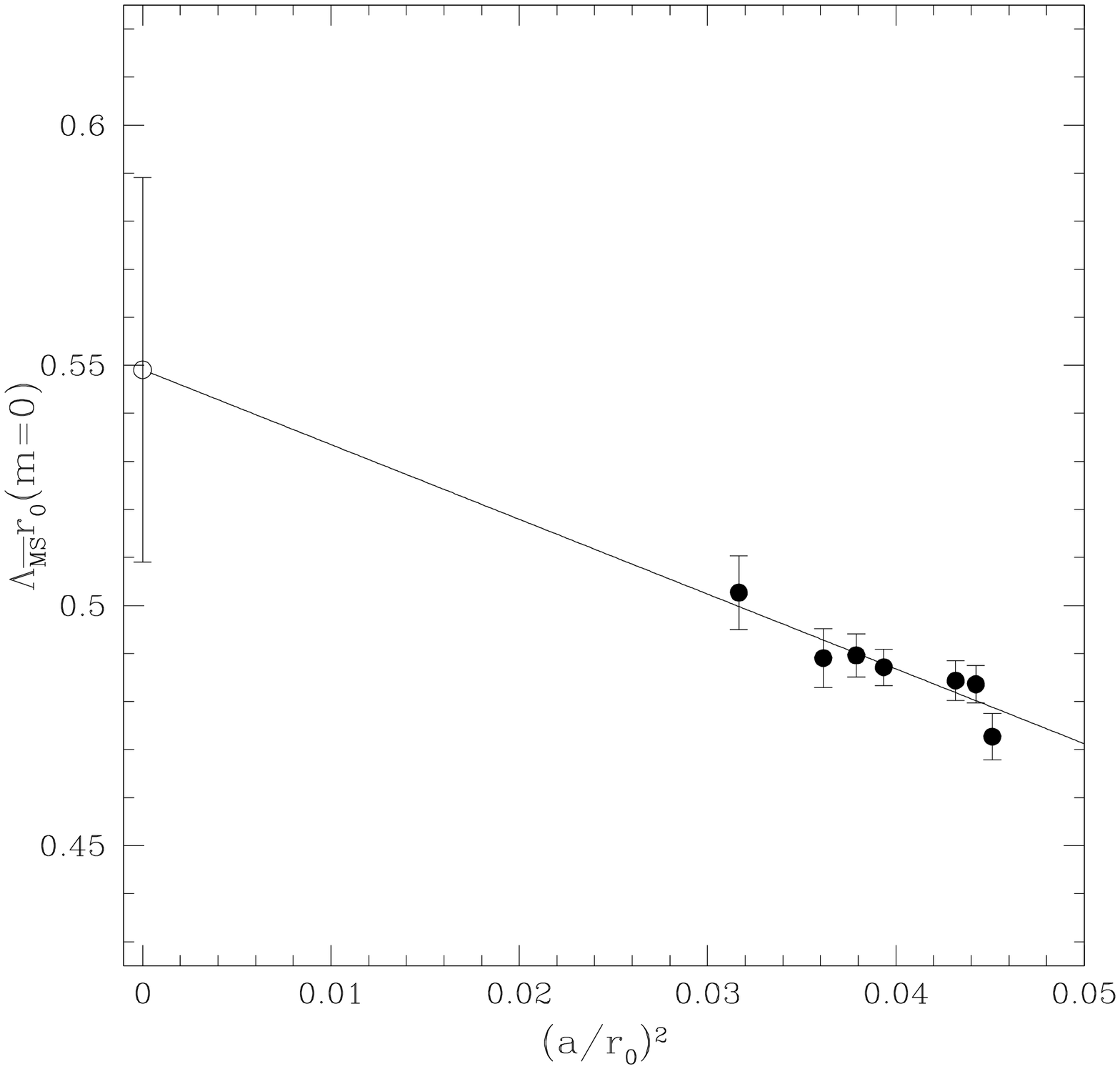,width=9.5cm}
\caption{The scale parameter $\Lambda_{\overline{MS}}\,r_0\,(m=0)$ against 
$(a/r_0)^2$ together 
with the fit (\ref{extra}) and the continuum result (\ref{contd}). The error 
of (\ref{contd}) shown is the statistical error only.}
\label{fig4}
\vspace{1.0cm}
\epsfig{file=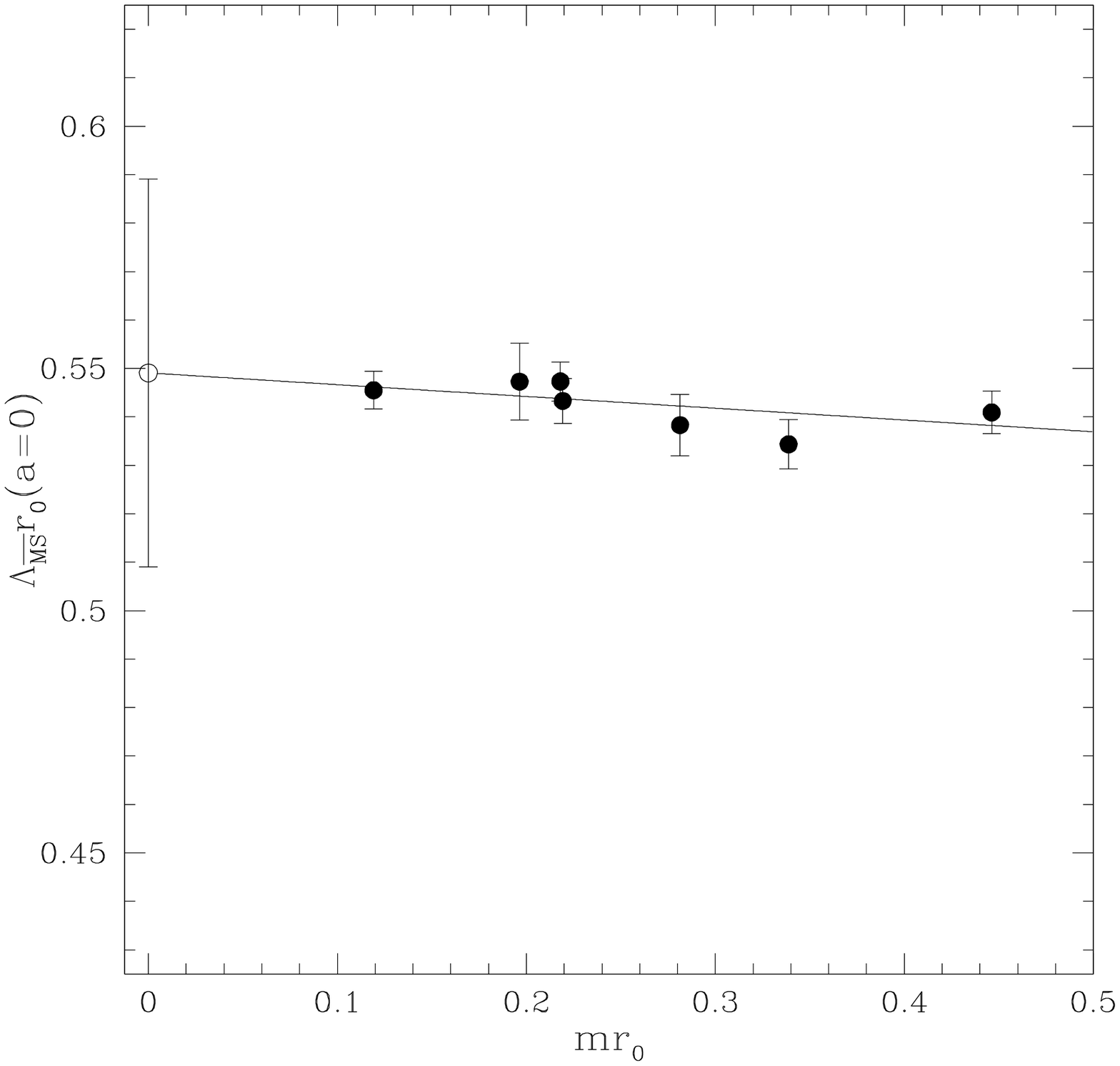,width=9.5cm}
\caption{The scale parameter $\Lambda_{\overline{MS}}\,r_0\,(a=0)$ against 
$m r_0$ together with 
the fit (\ref{extra}) and the continuum result (\ref{contd}).}
\label{fig5}
\end{center}
\end{figure}
\clearpage
\noindent


\vspace*{0.3cm}
\noindent
\underline{\it Comparison with Phenomenology}
\vspace*{0.5cm}

How can our results be compared with the phenomenological numbers? A fit 
to the world data of $\alpha_s$ gives the average value at the $Z$ 
mass~\cite{PDG} $\alpha_{\overline{MS}}^{N_f=5} (m_Z) = 0.118(2)$, which 
corresponds to $\Lambda_{\overline{MS}}^{N_f=5} = 208(25)\: \mbox{MeV}$. 
The latter value refers to an idealized world of five massless quarks and 
thus cannot be compared immediately to our numbers. We may 
extrapolate $\Lambda_{\overline{MS}}\, r_0$ to three flavors (remember that 
$r_0$ is extracted from the phenomenological heavy quark potential) and then 
evolve the corresponding $\alpha_{\overline{MS}}$ to the $Z$ mass, using the 
three-loop matching formulae~\cite{karlsruhe}. We do this by 
extrapolating  $\ln (\Lambda_{\overline{MS}}\, r_0 )$ 
linearly in $N_f$ to $N_f=3$, ignoring 
the fact that the strange quark mass is 
already relatively heavy and therefore less effective.
For $r_0 = 0.5 \: \mbox{fm}$ this gives $\Lambda_{\overline{MS}}^{N_f=3} = 
205(22)(20)\: \mbox{MeV}$. (Allowing for a 5\% uncertainty of the physical
scale parameter $r_0$ would increase the systematic error only slightly to 
$22\: \mbox{MeV}$.)
With the help of eq.~(\ref{soln}) we now compute 
$\alpha_{\overline{MS}}^{N_f=3}$ at the scale $\mu = 1 \: \mbox{GeV}$ and 
obtain $\alpha_{\overline{MS}}^{N_f=3} (1\, \mbox{GeV}) = 0.330(21)(19)$.
Taking the charm and bottom thresholds to be at 1.5 GeV and 4.5 GeV, 
respectively, we then find $\alpha_{\overline{MS}}^{N_f=5} (m_Z) = 
0.1076(20)(18)$, a number which is somewhat lower than the phenomenological 
value. If, on the other hand, we evolve the phenomenological value down to 
$N_f=4$ and $N_f=3$, we obtain $\Lambda_{\overline{MS}}^{N_f=4} = 
292(31)\:\mbox{MeV}$ and $\Lambda_{\overline{MS}}^{N_f=3} = 
342(32)\:\mbox{MeV}$, respectively. A logarithmic extrapolation to $N_f = 2$,
similar to our extrapolation of the lattice data but in reverse order, would 
give $\Lambda_{\overline{MS}}^{N_f=2} = 445(68)\:\mbox{MeV}$. 

In Fig.~6 we compare the $\Lambda$ values obtained by the various methods. 
At energy scales below the charm mass threshold the physics should be 
determined by $\Lambda_{\overline{MS}}^{N_f=3}$. So one would expect that the
lattice numbers extrapolate smoothly to the corresponding 
phenomenological value. We see, however, that this is not the case. 
The reason for this mismatch remains to be found. 

\begin{figure}[t]
\begin{center}
\vspace*{-0.7cm}
\epsfig{file=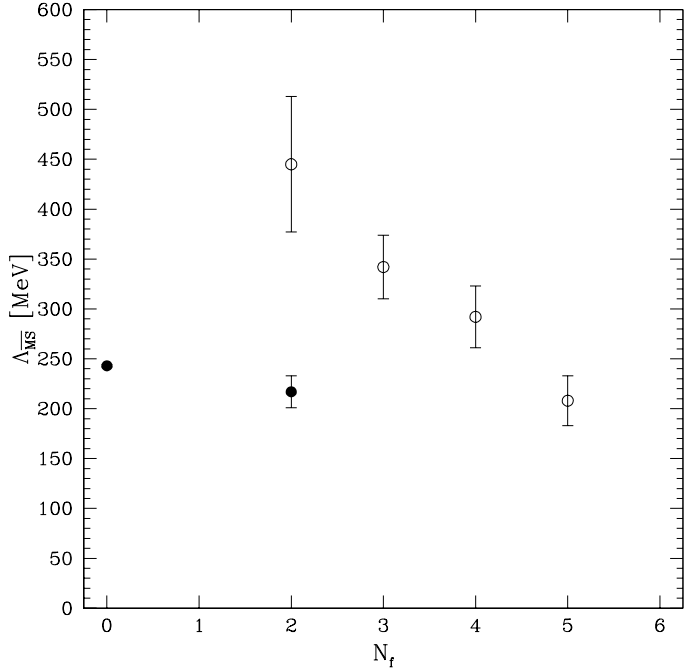,width=9.5cm}
\caption{The lattice (\lower0.85pt\hbox{\Large $\bullet$} ) and 
phenomenological (\lower0.85pt\hbox{\Large $\circ$} ) scale parameters 
$\Lambda_{\overline{MS}}$ against $N_f$. The error bars of the lattice
numbers correspond to the statistical errors.}
\label{fig6}
\end{center}
\end{figure}

\section{Conclusions}

Our quenched result agrees very well with results of other calculations 
using different methods. We find significant $O(a^2)$ corrections. For example 
at $\beta = 6.0$, corresponding to a lattice spacing $a \approx 0.1 \: 
\mbox{fm}$, they amount to $\approx 10\%$, which makes an extrapolation of the
results to the continuum limit indispensible. In the dynamical case the data 
cover a much smaller range of $a$, which makes the extrapolation to the 
continuum limit less reliable. But it is reassuring to see that the 
continuum limit is approached at a similar rate as in the quenched case. 

Our dynamical calculation is similar in spirit to previous unquenched
computations of $\alpha_s$~\cite{aoki,wingate,davies,sesam} (albeit not 
exactly the same). The main differences are that we are using a 
non-perturbatively $O(a)$ improved fermion action, which reduces cut-off 
effects, the conversion to $g_{\overline{MS}}$ is done consistently
in two-loop perturbation theory, and an extrapolation of $\Lambda$ to 
the chiral and continuum limit is performed.

\section*{Appendix}

We follow the argument in~\cite{MM} calculating the relation between the
$\Lambda$ parameters from the potential. We require that the potential 
(or force) between two static charges should be the same, whether computed 
as a series in $g^2$ or $g^2_{\overline{MS}}$.   
All we have to do to calculate the fermionic piece of the relation 
is to compute the fermionic contribution to the gluon propagator. 

In the scheme $\mathcal{S}$ we have 
\begin{equation}
\begin{split}
V(\vec{r}) &= \int \frac{\mbox{d}^4 q}{(2 \pi)^4} 2 \pi \delta(q_4) 
\frac{g_{\mathcal{S}}^2}{q^2} \big(1 - g_{\mathcal{S}}^2 G^{\mathcal{S}}(q^2)\\
&- g_{\mathcal{S}}^2 N_f \Pi^{\mathcal{S}}(q^2,m) +O(g_{\mathcal{S}}^4)\big) 
\mbox{e}^{{\rm i} \vec{q}\,\vec{r}} , 
\end{split}
\end{equation} 
where $G^{\mathcal{S}}(q^2)$ is the one-loop gluon contribution to the 
potential, and $\Pi^{\mathcal{S}}$ is the one-loop quark vacuum polarization.  
If two schemes, $\mathcal{S}$ and $\mathcal{S}'$, are to give the same answer 
for $V(\vec{r})$, their couplings have to be related by
\begin{equation}
\begin{split}
\frac{1}{g_{\mathcal{S}}^2} = \frac{1}{g_{\mathcal{S}'}^2} 
&+\big( G^{\mathcal{S}'}(q^2) - G^{\mathcal{S}}(q^2) \big) 
+ N_f \big( \Pi^{\mathcal{S}'}(q^2,m) \\
&- \Pi^{\mathcal{S}}(q^2,m) \big) 
+O(g^2) .
\end{split}
\end{equation} 
To find the fermionic part of the conversion from $g^2$ to 
$g^2_{\overline{MS}}$, we have to calculate the vacuum polarization in both 
schemes and take the difference. 

In the $\overline{MS}$ scheme we find
\begin{equation}
\Pi^{\overline{MS}}(q^2,m) = 
-\frac{1}{24 \pi^2} \big( \ln (q^2/\mu^2) - \frac{5}{3} 
+ O(m^2/q^2) \big) . 
\label{Pimsb} 
\end{equation} 
On the lattice we obtain
\begin{equation}
\begin{split}
\Pi(q^2,m) &= -\frac{1}{24 \pi^2} 
\Big( \ln (a^2 q^2) -3 a m (1 - c_{SW}) \ln (a^2 q^2)  \\
&- 3.25275141(5)  +1.19541770(1) c_{SW} - 7.06903716(4) c_{SW}^2 \\
&+ a m  \big( 6.46270704(30) -  5.29413266(6) c_{SW} \\ 
&+1.67389761(2)  c_{SW}^2 \big) \Big) .
\label{Pi_lat}
\end{split}
\end{equation}
We see that there is an unwanted $a m \ln (a^2 q^2)$ term 
in eq.~(\ref{Pi_lat}) unless $c_{SW} = 1 + O(g^2)$. Combining 
our calculation of $\Pi$ with the calculation of the purely 
gluonic part in the literature~\cite{Luscher}, we get our final result 
for $t_1$ (for general $N_c$): 
\begin{equation}
\begin{split} 
t_1 &= 0.16995600 \,N_c -\frac{1}{8 N_c} \\ 
&- N_f \big( 0.00669600 - 0.00504671\, c_{SW} +  0.02984347\, c_{SW}^2  \\ 
&+ a m ( -0.02728371 + 0.02235032\, c_{SW} - 0.00706672\, c_{SW}^2 )\big) .
\label{t_dyn2}
\end{split}
\end{equation}

The $a m$ term in eq.~(\ref{t_dyn2}) means that in dynamical QCD the contours 
of constant $g_{\overline{MS}}^2(1/a)$ will be slanted when plotted in a 
($a m, \beta$) plane. As one expects the contours of constant $r_0/a$ to 
roughly follow contours of constant $g_{\overline{MS}}^2(1/a)$, this term 
gives a possible explanation of the appearance of the sloped lines in Fig.~3.

\section*{Acknowledgement}

We thank H. Panagopoulos for communicating the three-loop $\beta$ function 
for improved Wilson fermions to us prior to publication.
The numerical calculations have been performed on the Hitachi SR8000 at LRZ
(Munich), on the Cray T3E at EPCC (Edinburgh), NIC (J\"ulich) and ZIB (Berlin)
as well as on the APE/Quadrics at DESY (Zeuthen). We thank all institutions
for their support. This work has been supported in part by the European 
Community's Human Potential Program under contract HPRN-CT-2000-00145, 
Hadrons/Lattice QCD. MG and PELR acknowledge financial
support from DFG.

\end{document}